\documentclass[sigconf]{acmart}
\usepackage{multirow} 
\usepackage{adjustbox}
\usepackage{bm}
\usepackage{lipsum}
\usepackage{xspace}
\usepackage{cleveref}
\usepackage{pgfplots}
\usepackage{subfig} 
\usepackage{xcolor}
\usepackage{geometry}
\usepackage{pgfplots}
\usepackage{enumitem}
\usepackage{tcolorbox}

\AtBeginDocument{%
  }

\copyrightyear{2025}
\acmYear{2025}
\setcopyright{rightsretained}
\acmConference[CIKM '25] {Proceedings of the 34th ACM International Conference on Information and Knowledge Management}{ November 10--14, 2025}{Seoul, Republic of Korea.}
\acmBooktitle{Proceedings of the 34th ACM International Conference on Information and Knowledge Management (CIKM '25), November 10--14, 2025, Seoul, Republic of Korea}
\acmISBN{979-8-4007-2040-6/2025/11}
\acmDOI{10.1145/XXXXXX.XXXXXX}

\begin{document}

\title[Do RecSys Really Leverage Multimodal Content? A Comprehensive Analysis on Multimodal Representations]{Do Recommender Systems Really Leverage Multimodal Content? A Comprehensive Analysis on Multimodal\\ Representations for Recommendation}

\author{Claudio Pomo}
\authornote{Corresponding authors: Claudio Pomo (\url{claudio.pomo@poliba.it}), Matteo Attimonelli (\url{matteo.attimonelli@poliba.it}),
and Danilo Danese (\url{danilo.danese@poliba.it}).}
\orcid{0000-0001-5206-3909}
\email{claudio.pomo@poliba.it}
\affiliation{%
  \institution{Politecnico Di Bari}
  \city{Bari}
  \country{Italy}}

\author{Matteo Attimonelli}
\authornotemark[1]
\orcid{0009-0003-6600-1938}
\email{matteo.attimonelli@poliba.it}
\affiliation{%
  \institution{Politecnico Di Bari}
  \city{Bari}
  \country{Italy}}
\affiliation{%
  \institution{Sapienza University of Rome}
  \city{Rome}
  \country{Italy}
}

\author{Danilo Danese}
\authornotemark[1]
\orcid{0009-0000-5203-1229}
\email{danilo.danese@poliba.it}
\affiliation{%
  \institution{Politecnico Di Bari}
  \city{Bari}
  \country{Italy}
}

\author{Fedelucio Narducci}
\orcid{0000-0002-9255-3256}
\email{fedelucio.narducci@poliba.it}
\affiliation{%
  \institution{Politecnico Di Bari}
  \city{Bari}
  \country{Italy}}

\author{Tommaso Di Noia}
\orcid{0000-0002-0939-5462}
\email{tommaso.dinoia@poliba.it}
\affiliation{%
  \institution{Politecnico Di Bari}
  \city{Bari}
  \country{Italy}}

\newcommand{\matteo}[1]{\textcolor{red}{{\bf [Matteo: }{\em #1}{\bf ]}}}
\newcommand{\cla}[1]{\textcolor{blue}{{\bf [Cla: }{\em #1}{\bf ]}}}
\newcommand{\danilo}[1]{\textcolor{purple}{{\bf [Danilo: }{\em #1}{\bf ]}}}
\newcommand{\qwen}{\textsc{Qwen2-VL}\xspace}
\newcommand{\phiv}{\textsc{Phi-3.5-VI}\xspace}
\newcommand{\rnet}{\textsc{RNet50}\xspace}
\newcommand{\vit}{\textsc{ViT}\xspace}
\newcommand{\sbert}{\textsc{Sbert}\xspace}
\newcommand{\clip}{CLIP\xspace}
\newcommand{\rnetext}{\textsc{ResNet50}\xspace}
\newcommand{\sbertext}{\textsc{Sentence-Bert}\xspace}

\definecolor{tab10blue}{HTML}{1f77b4}
\definecolor{tab10orange}{HTML}{ff7f0e}
\definecolor{tab10green}{HTML}{2ca02c}
\definecolor{tab10red}{HTML}{d62728}
\definecolor{tab10purple}{HTML}{9467bd}
\definecolor{tab10brown}{HTML}{8c564b}
\definecolor{tab10pink}{HTML}{e377c2}
\definecolor{tab10gray}{HTML}{7f7f7f}
\definecolor{tab10olive}{HTML}{bcbd22}
\definecolor{tab10cyan}{HTML}{17becf}
\definecolor{color1}{HTML}{9ACBD0}
\definecolor{color2}{HTML}{48A6A7}
\definecolor{color3}{HTML}{006A71}
\renewcommand{\shortauthors}{Pomo et al.}

\begin{abstract}
Multimodal Recommender Systems aim to improve recommendation accuracy by integrating heterogeneous content, such as images and textual metadata. While effective, it remains unclear whether their gains stem from true multimodal understanding or increased model complexity. This work investigates the role of multimodal item embeddings, emphasizing the semantic informativeness of the representations. Initial experiments reveal that embeddings from standard extractors (e.g., \rnetext, \sbertext) enhance performance, but rely on modality-specific encoders and ad hoc fusion strategies that lack control over cross-modal alignment. To overcome these limitations, we leverage Large Vision–Language Models (LVLMs) to generate \textit{multimodal-by-design} embeddings via structured prompts. This approach yields semantically aligned representations without requiring any fusion. Experiments across multiple settings show notable performance improvements.
Furthermore, LVLMs embeddings offer a distinctive advantage: they can be decoded into structured textual descriptions, enabling direct assessment of their multimodal comprehension. When such descriptions are incorporated as side content into recommender systems, they improve recommendation performance, empirically validating the semantic depth and alignment encoded within LVLMs outputs. Our study highlights the importance of semantically rich representations and positions LVLMs as a compelling foundation for building robust and meaningful multimodal representations in recommendation tasks.


\end{abstract}

\ccsdesc[500]{Computing methodologies~Natural language processing}
\ccsdesc[500]{Computing methodologies~Information extraction}
\ccsdesc[300]{Computing methodologies~Learning latent representations}
\ccsdesc[500]{Information systems~Recommender systems}
\ccsdesc[500]{Information systems~Information extraction}




\maketitle

\section{Introduction}
Recommender Systems (RSs) have emerged as one of the most extensively studied and widely applied technologies, facilitating personalized and on-demand information services. Despite the success of collaborative filtering techniques in large-scale platforms, RSs continue to encounter inherent limitations, including data sparsity~\cite{DBLP:conf/www/SarwarKKR01,DBLP:conf/icdm/HuKV08}, cold-start~\cite{DBLP:conf/sigir/ScheinPUP02,DBLP:conf/iui/RashidACLMKR02} problems, scalability constraints, which ultimately hinder recommendation accuracy. In response to these challenges, \textit{Multimodal Recommender Systems} (MMRSs)~\cite{DBLP:conf/aaai/HeM16,DBLP:conf/mm/Zhang00WWW21,DBLP:conf/www/ZhouZLZMWYJ23,DBLP:conf/mm/ZhouS23} have become a cutting-edge solution for enhancing user satisfaction and engagement. By integrating multiple content modalities, such as images and textual descriptions of items, MMRSs improve the recommendation pipeline, addressing data sparsity issues and enabling more precise and tailored recommendations.
While MMRSs have demonstrated empirical success in leveraging multimodal content to boost accuracy~\cite{DBLP:journals/tors/MalitestaCPMNS25}, a critical question remains insufficiently addressed: \textit{Do MMRSs truly benefit from the semantic content of input features, or are observed performance gains mainly related to increased model complexity?}

This work addresses the aforementioned question through a comprehensive investigation of multimodal embeddings in recommender systems. We begin by examining whether the performance gains reported in MMRSs arise from genuine cross-modal understanding or are instead attributable to incidental factors such as increased model capacity (e.g., number of parameters). 
To this end, we compare classical collaborative filtering models with various multimodal alternatives through a series of controlled experiments, including the use of (a) Gaussian noise, (b) multivariate structured noise, and (c) multimodal item embeddings. Our findings confirm that multimodal content can indeed enhance recommendation performance relative to classical baselines. Furthermore, we note critical limitations in how multimodal features are typically integrated into MMRSs:
most existing systems rely on pre-trained encoders, such as~\rnetext~\cite{DBLP:conf/cvpr/HeZRS16} for visual content and~\sbertext~\cite{DBLP:conf/emnlp/ReimersG19} for textual metadata, as they represent the most widely adopted feature extractors in the field.
The embeddings from these encoders are combined using methods like concatenation, summation, or element-wise multiplication based on dimension compatibility~\cite{DBLP:journals/tors/MalitestaCPMNS25, DBLP:conf/mm/ZhouS23, DBLP:conf/mm/Zhang00WWW21, DBLP:conf/www/ZhouZLZMWYJ23}. This straightforward fusion approach has two main issues. First, it lacks a systematic way to manage the weighting and preservation of information from each modality. Second, it offers limited visibility into how recommender systems use the resulting composite representations, reducing control over the distribution of modality-specific information. Consequently, the integration of multimodal signals may lack transparency and robustness, thus undermining the semantic richness intended by multimodal data.

Motivated by these findings, we explore a more principled alternative: the use of \textit{multimodal-by-design} representations, which are embeddings generated by models trained to natively process and align multiple modalities. In particular, we focus on Large Vision–Language Models (LVLMs)~\cite{DBLP:conf/nips/LiuLWL23a,DBLP:conf/nips/Dai0LTZW0FH23}, which extend Large Language Models (LLMs)~\cite{DBLP:journals/corr/abs-2302-13971,DBLP:conf/nips/BrownMRSKDNSSAA20} by incorporating advanced computer vision capabilities. These models are trained to unify visual and textual information within a shared semantic space, effectively tackling tasks such as image captioning and visual question answering (VQA)~\cite{DBLP:conf/iccv/AntolALMBZP15,DBLP:conf/cvpr/VinyalsTBE15}.
By prompting LVLMs to describe the visual content of item images, we facilitate embedding extraction through a VQA task, effectively leveraging their in-context learning (ICL) capability (e.g., the ability to adapt to new tasks without explicit fine-tuning). 
The resulting embeddings capture cross-modal semantics by coherently integrating visual and linguistic information within a unified latent space. Unlike traditional late-fusion methods that combine separately encoded unimodal features, our experiments demonstrate that this approach yields more effective representations, leading to improved MMRS performance.

Furthermore, differently from traditional late-fusion approaches, embeddings extracted from LVLMs can be decoded into the original structured textual descriptions.
This allows the designer to directly inspect the multimodal understanding of the LVLMs through analysis of the generated outputs. To evaluate the semantic informativeness of these textual representations and the associated multimodal-by-design embeddings, we incorporate the generated descriptions as additional content signals in collaborative recommendation models. The resulting improvements in recommendation performance, measured against both simple collaborative filtering baselines and more advanced MMRSs, serve as indicators of how well these descriptions capture the relevant item semantics and their practical utility in recommendation. Experimental results show that leveraging such descriptions not only enhances recommendation quality relative to classical models but, in some cases, matches or surpasses the performance of more complex MMRS architectures. This confirms the strong multimodal comprehension of LVLMs and the rich semantic content embedded within their latent representations.

To summarize, our contributions are the following:
\begin{itemize}
    \item We demonstrate that the performance of MMRSs is primarily influenced by the quality and informativeness of the input multimodal features, rather than merely by the model capacity (e.g., number of parameters).
 
    \item We investigate the potential of state-of-the-art LVLMs as a principled alternative for generating multimodal item embeddings, given their ability to natively integrate visual and textual modalities without requiring explicit fusion of separately encoded features, thus preserving transparency and robustness of the recommendation results.
    
    \item We provide empirical evidence that embeddings extracted from LVLMs capture the semantic content of items by analyzing the textual descriptions they generate and directly associate with item embeddings. Moreover, incorporating these descriptions as side content into classical recommender models enhances performance, achieving results comparable to MMRSs.
\end{itemize}


The remainder of the paper reviews relevant literature on multimodal recommendation and vision–language modeling, details our methodology and experimental setup, presents the results, and concludes with directions for future work.
\section{Related Work}

This section reviews the key areas supporting our study: MMRSs and LVLMs. We first examine MMRSs, highlighting challenges in achieving true cross-modal understanding, particularly with ad hoc fusion of unimodal features. We then turn to LVLMs as a promising paradigm for unified multimodal representation learning. Together, these viewpoints contextualize our inquiry into the potential of LVLMs to provide a more efficacious and theoretically grounded approach to multimodal recommendation.

\subsection{Multimodal Recommendation}
MMRSs aim to enrich the semantics of user and item representations by integrating item-side multimodal content, such as textual descriptions and visual illustrations, with traditional collaborative filtering signals~\cite{DBLP:conf/mm/LiuYLWTZSM21}. This integration is particularly valuable for addressing common issues like data sparsity and the cold-start problem~\cite{DBLP:journals/tois/LiuXGWLH23, DBLP:journals/tors/MalitestaCPMNS25}, leading to improved recommendation accuracy across diverse domains including fashion~\cite{DBLP:conf/sigir/ChenCXZ0QZ19}, music~\cite{DBLP:conf/recsys/OramasNSS17}, and micro-videos~\cite{DBLP:journals/tmm/CaiQFX22}. Early approaches, such as VBPR~\cite{DBLP:conf/aaai/HeM16}, employed matrix factorization to combine multimodal information with ID embeddings. With the advancement of deep learning, more sophisticated techniques have been adopted, including variational autoencoders~\cite{DBLP:conf/kdd/TranL22,DBLP:conf/kdd/ZhangYLXM16} and graph neural networks (GNNs), e.g. FREEDOM~\cite{DBLP:conf/mm/ZhouS23} and LATTICE~\cite{DBLP:conf/mm/Zhang00WWW21}, with notable early work by~\citet{DBLP:conf/mm/WeiWN0HC19} utilizing GNNs for modality representation learning and fusion. Self-supervised learning methods have also been utilized to model relationships between different modalities~\cite{DBLP:journals/tmm/TaoLXWYHC23,DBLP:conf/www/ZhouZLZMWYJ23}, and meta-learning approaches like MML~\cite{DBLP:conf/cikm/PanCTLW0Z22} have been applied to tackle cold-start issues using multimodal side information. Furthermore, attention mechanisms can enhance the interpretability of recommendations~\cite{DBLP:journals/kais/FangLS25}. Despite these advancements, challenges persist, particularly in learning effectively from limited or potentially biased and noisy user-item interaction data~\cite{DBLP:conf/sigir/ChenCXZ0QZ19, DBLP:conf/mm/YangFZWL23, DBLP:conf/mm/LiuTSYH22} and in achieving efficient fusion of diverse multimodal knowledge~\cite{DBLP:journals/tmm/LiuCCLNK23,DBLP:journals/tois/LiuWLWNC24,DBLP:journals/tkde/ZhangZLZWW23,DBLP:journals/tors/MalitestaCPMNS25}. 
Traditional multimodal approaches often rely on late-fusion techniques, such as concatenation, summation, or element-wise multiplication, to combine features from separate unimodal encoders~\cite{DBLP:conf/mm/ZhouS23, DBLP:conf/mm/Zhang00WWW21, DBLP:conf/www/ZhouZLZMWYJ23}. However, these methods lack explicit mechanisms for controlling information flow and offer limited transparency into how downstream models utilize the resulting composite representations. This raises questions about whether performance gains reflect genuine semantic understanding or merely increased model capacity. Additionally, it remains unclear whether such fusion strategies effectively preserve and integrate cross-modal information. These limitations highlight a critical gap in the literature that this work aims to address.


\sloppy
\subsection{Large Vision-Language Models}
Building on the capabilities of LLMs~\cite{DBLP:conf/emnlp/Dong0DZMLXX0C0S24, DBLP:conf/nips/BrownMRSKDNSSAA20, DBLP:conf/iclr/PatelLRCRC23}, Large Vision-Language Models (LVLMs)~\cite{DBLP:journals/corr/abs-2405-02246, DBLP:conf/nips/LiuLWL23a, DBLP:journals/corr/abs-2404-01331, DBLP:journals/corr/abs-2402-11530, DBLP:conf/nips/Dai0LTZW0FH23, DBLP:journals/tmlr/WangYHLLGLLW22, DBLP:journals/corr/abs-2404-14219, DBLP:journals/corr/abs-2407-21783, DBLP:journals/corr/abs-2409-12191} extend this paradigm by integrating computer vision capabilities, enabling joint reasoning over textual and visual inputs. This multimodal integration makes LVLMs well-suited for tasks such as image captioning~\cite{DBLP:conf/cvpr/VinyalsTBE15} and visual question answering (VQA)~\cite{DBLP:conf/iccv/AntolALMBZP15}.
These tasks fundamentally involve verbalizing visual content, either by generating descriptive text or answering queries based on image understanding, and require sophisticated alignment and interpretation of visual and textual signals to achieve a deep semantic understanding of multimodal data. This ability to process and understand multimodal information makes LVLMs particularly relevant for domains such as MMRSs.
A critical capability that underpins the versatility of both LLMs and LVLMs is In-Context Learning (ICL)~\cite{DBLP:conf/emnlp/Dong0DZMLXX0C0S24, DBLP:conf/nips/BrownMRSKDNSSAA20, DBLP:conf/iclr/PatelLRCRC23}. ICL allows models to adapt to new tasks by leveraging a few illustrative examples provided directly within the input prompt, avoiding the need for computationally expensive retraining or fine-tuning. Specifically, Few-Shot ICL~\cite{DBLP:conf/iclr/PatelLRCRC23, DBLP:conf/nips/BrownMRSKDNSSAA20} enables effective generalization with only a limited number of labeled instances, which is highly advantageous in data-scarce scenarios.
Leveraging these advancements, our work explores the application of LVLMs embeddings within MMRSs. 
While prior work has explored multimodal integration in recommender systems using various strategies~\cite{DBLP:journals/tors/MalitestaCPMNS25, DBLP:conf/www/AttimonelliDMPG24}, to the best of our knowledge, this is the first systematic use of LVLMs embeddings for this purpose. Unlike traditional approaches that fuse features from separate unimodal encoders, LVLMs features are inherently multimodal (\textit{multimodal-by-design}~\cite{attimonelli2024ducho, DBLP:conf/www/AttimonelliDMPG24}), as they are learned from models jointly trained across modalities, capturing cross-modal interactions natively. Moreover, most of the LVLMs research in recommendation has focused on textual augmentation~\cite{10825030}, rather than directly leveraging the rich multimodal embeddings. 
Our approach thus builds upon and contributes to the intersection of Multimodal and Recommender Systems research.

\section{Preliminaries}
\label{sec:methodology}
This work investigates the task of personalized recommendation in both classical and multimodal settings. The goal is to predict user preferences for unseen items based on prior interactions and, when available, multimodal content. In the classical setting, the task is formalized as the learning of a function:
\begin{equation}
    f: \mathcal{U} \times \mathcal{I} \rightarrow \mathbb{R},
\end{equation}
which predicts a relevance score $\hat{y}_{ui} \in \mathbb{R}$ expressing the likelihood that user $u \in \mathcal{U}$ will prefer item $i \in \mathcal{I}$. Users and items are encoded as latent vectors $\mathbf{z}_u, \mathbf{z}_i \in \mathbb{R}^d$, where $d$ denotes the dimensionality of the latent space. The relevance is computed via a scoring function $\phi: \mathbb{R}^d \times \mathbb{R}^d \rightarrow \mathbb{R}$:
\begin{equation}
    \hat{y}_{ui} = \phi(\mathbf{z}_u, \mathbf{z}_i),
\end{equation}
where $\phi$ can be a dot product, cosine similarity, or a parameterized function. User embeddings may be directly learned or constructed by aggregating the embeddings of previously interacted items:
\begin{equation}
    \mathbf{z}_u = g(\{\mathbf{z}_{i_j} \mid i_j \in \mathcal{I}_u\}),
    \label{eq:mmrs}
\end{equation}
with $\mathcal{I}_u \subseteq \mathcal{I}$ denoting the set of items interacted with user $u$, and $g(\cdot)$ being an aggregation function.

This framework is extended to a multimodal setting in which each item $i \in \mathcal{I}$ is associated with a set of modality-specific features $\{\mathbf{x}_i^{(m)} \in \mathbb{R}^{d_m} \mid m \in \mathcal{M}\}$, extracted from pre-trained models. Each modality $m \in \mathcal{M}$ is processed by a corresponding encoder $h^{(m)}: \mathbb{R}^{d_m} \rightarrow \mathbb{R}^d$, resulting in
\begin{equation}
    \mathbf{z}_i^{(m)} = h^{(m)}(\mathbf{x}_i^{(m)}).
\end{equation}
The item representation is enriched by the available modalities, yielding a content-aware recommendation function:
\begin{equation}
    f(u, \{\mathbf{x}_i^{(m)}\}_{m \in \mathcal{M}}) = \phi(\mathbf{z}_u, \mathbf{z}_i) = \hat{y}_{ui}
    \label{eq:mm_rec}
\end{equation}
where $\mathbf{z}_i$ integrates the multimodal features.

To investigate the role of multimodal representations, embeddings from three categories of feature extractors are considered: classical unimodal encoders, general-purpose multimodal encoders, and LVLMs. For classical models, features are extracted either from the penultimate layer or, in the case of transformers-based encoders, from the \texttt{[CLS]} (short for "Classification") token, which is a special classification token prepended to the input sequence whose final hidden state is typically used as a summary representation of the entire input~\cite{DBLP:conf/naacl/DevlinCLT19}. In recommendation context, visual representations are derived from product images, while textual representations are obtained from item descriptions. For general-purpose multimodal encoders, embeddings are taken from the shared projection layer that aligns information across modalities into a unified latent space.


Differently from multimodal encoders, LVLMs are primarily designed for generative tasks such as VQA, rather than for embedding extraction. To adapt these models for use in recommendation, embedding extraction is reformulated as a VQA task: each item image is paired with a structured query designed to find salient attributes. This formulation enables the application of ICL~\cite{DBLP:conf/nips/BrownMRSKDNSSAA20, DBLP:conf/iclr/PatelLRCRC23}, a paradigm in which models generalize to new tasks by conditioning on a small set of illustrative examples embedded within the input prompt, thereby avoiding the need for task-specific fine-tuning. Using one-shot prompting, where a single example response is provided, LVLMs are guided to generate semantically informative textual descriptions (see~\Cref{subsec:lvlm_preprocessing}). These outputs function as content representations suitable for recommendation, while also providing access to the internal states of the model that reflect its multimodal semantic understanding.


Due to the autoregressive nature of LVLMs, embeddings are extracted from the final hidden state associated with the last output token, specifically, the \texttt{[EOS]} (End-of-Sequence) token. This special token denotes the termination of a sequence and is commonly used in autoregressive models to signal the point at which text generation should stop. The hidden state corresponding to the \texttt{[EOS]} token captures the model’s aggregated understanding of the entire input context and its expected response, providing a semantically aligned representation. Unlike mean pooling~\cite{DBLP:conf/iclr/Lee0XRSCP25}, which averages token embeddings across a sequence without regard to their order or contextual role, using the \texttt{[EOS]} representation~\cite{DBLP:conf/iclr/LiQXCLLSL25, DBLP:conf/acl/WangYHYMW24} preserves the model’s causal structure and is suitable for downstream recommendation tasks.
The resulting embeddings are inherently multimodal, eliminating the need for late-fusion of unimodal features. Additionally, to assess the multimodal comprehension of LVLMs, their generated textual representations can be integrated into classical recommenders as auxiliary side information, enabling a quantitative evaluation of their informativeness. This dual use of LVLMs outputs, as both latent embeddings and interpretable text, supports a comprehensive assessment of their semantic relevance and utility in recommendation.

\section{Experimental Setting}
\label{sec:experimental_setting}







This section outlines the experimental setup for validating the research questions. It begins with a description of the dataset used in the evaluation, followed by the multimodal feature extractors for item representations. Next, we introduce the trained recommendation models, and finally, we explain how to process outputs from LVLMs into content features for recommendation tasks.

\subsection{Datasets}

For this study, we selected three distinct product categories, Baby, Pets, and Clothing, from the comprehensive Amazon Reviews 2023 dataset~\cite{DBLP:journals/corr/abs-2403-03952}. This dataset provides extensive user-item interactions and rich multimodal metadata, including both visual and textual descriptions.
These particular categories were chosen primarily because they exhibit inherent differences in the distribution and interaction patterns between users and items within the larger dataset, providing distinct characteristics for our analysis.
Subsequently, specific preprocessing steps were applied to each of these selected category subsets. To ensure a minimum level of user and item activity within each set, k-core filtering was performed: the Baby and Pets categories were processed as 5-core, requiring every user and item to have at least 5 associated reviews. The Clothing category was processed as a 10-core, needing at least 10 reviews per user and item.
Furthermore, to align with our focus on multimodal information, we exclusively included items that possessed both an associated textual description and at least one accompanying image. In cases where multiple images were available for a selected item, the first image identified by the \texttt{large} key within the metadata was consistently utilized.
Descriptive statistics summarizing these curated category datasets are presented in~\Cref{tab:datasetInfo}.
Finally, each processed category was partitioned into standard training, validation, and test sets following an 80\%/10\%/10\% split, respectively, to facilitate model development and evaluation.

\subsection{Multimodal Feature Extractors}
\label{subsec:mm_feat_extractors}
To evaluate the extent to which MMRSs effectively leverage the semantic content of multimodal inputs, we conduct experiments using a diverse set of pre-trained feature extractors spanning visual and textual modalities. In addition to these learned representations, we also include synthetic noisy baselines generated from random noise and multivariate Gaussian distributions to assess model sensitivity to semantically uninformative features.
We selected established encoders to perform unimodal feature extraction. For visual features, we employ \textbf{\rnetext (\rnet)}~\cite{DBLP:conf/cvpr/HeZRS16}, a widely-used convolutional neural network, from which we extract activations from the final pooling layer as image embeddings. We also utilize the \textbf{Vision Transformer (\vit)}~\cite{DBLP:conf/iclr/DosovitskiyB0WZ21}, a more recent transformer-based architecture for image understanding that models spatial relationships using self-attention. For \vit, item representations are derived from the output embedding of the \texttt{[CLS]} token in the final transformer layer, specifically using the \textit{vit-base-patch16-224} checkpoints.
For textual content, we use \textbf{\sbertext (\sbert)}~\cite{DBLP:conf/emnlp/ReimersG19}, a sentence-level textual encoder built on BERT and fine-tuned for semantic textual similarity. \sbert maps item descriptions to dense vector representations via mean pooling over token embeddings, and we considered the \textit{all-mpnet-base-v2} checkpoints. 

Furthermore, to represent classical multimodal models trained with explicit vision-language alignment, we include \textbf{CLIP}~\cite{DBLP:conf/icml/RadfordKHRGASAM21}. This model is trained via contrastive learning to embed images and text into a shared latent space, and we extract the final projection from both the image and text encoders onto this shared space, utilizing the \textit{clip-vit-large-patch14} checkpoints.

To incorporate more recent, inherently multimodal-by-design representations, we employed two LVLMs. Our choice was guided by their performance on general vision-language benchmarks~\cite{DBLP:journals/corr/abs-2404-14219, DBLP:journals/corr/abs-2409-12191, DBLP:journals/corr/abs-2409-02813}, their proficient instruction-following capabilities, which are crucial for our VQA-based embedding extraction methodology (detailed in Section~\ref{subsec:lvlm_preprocessing}), and their open availability at the time of our research. These models, \textbf{Phi-3.5-vision-instruct (\phiv)}~\cite{DBLP:journals/corr/abs-2404-14219} and \textbf{\qwen-7B-Instruct (\qwen)}~\cite{DBLP:journals/corr/abs-2409-12191}, offer a practical balance of advanced capabilities and computational cost for our experiments. Specifically,~\phiv is a lightweight multimodal autoregressive model designed to process and generate coherent image-grounded textual responses. For each item, we input the image along with a prompt instructing the model to describe its visual content.
We then extract the final token embedding (i.e., the \texttt{[EOS]} representation) from the last hidden layer, which captures the semantics of both the input and the generated description by leveraging the autoregressive structure inherent to LVLMs.
We consider the \textit{Phi-3.5-vision-instruct} checkpoints.
Similarly, \qwen is a vision–language model capable of complex cross-modal reasoning. Following a comparable protocol to that used for~\phiv, we provide both the image and a natural language prompt, and extract the \texttt{[EOS]} token embedding from the final hidden layer, relying on the \textit{\qwen-7B-Instruct} checkpoints. 
These selected models span a range of architectural paradigms, offering a comprehensive suite for analyzing how the nature and quality of multimodal representations influence downstream recommendation performance.

\begin{figure*}[t!]
    \centering
    \small
    \begin{tcolorbox}[width=\linewidth, colback=gray!5, colframe=gray, boxrule=0.5pt, title=Prompt for Baby Products]
    Imagine you’re creating metadata for an image database of baby products. Your task is to select five keywords that best represent the image content. Fill in the blanks: \textbf{[Category]}, \textbf{[Age Group]}, \textbf{[Purpose]}, \textbf{[Material]}, \textbf{[Usage Context]}.\\[0.5em]
    For example, your answer will be: [Category] \{Feeding\}, [Age Group] \{Infant\}, [Purpose] \{Hygiene\}, [Material] \{Silicone\}, [Usage Context] \{Home\}.
    \end{tcolorbox}
    
    \begin{tcolorbox}[width=\linewidth, colback=gray!5, colframe=gray, boxrule=0.5pt, title=Prompt for Pets Items]
    Imagine you’re creating metadata for an image database of pet-related items. Your task is to select five keywords that best represent the image content. Fill in the blanks: \textbf{[Category]}, \textbf{[Pet Type]}, \textbf{[Purpose]}, \textbf{[Material]}, \textbf{[Usage Context]}.\\[0.5em]
    For example, your answer will be:  [Category] \{Toys\}, [Pet Type] \{Dog\}, [Purpose] \{Entertainment\}, [Material] \{Rubber\}, [Usage Context] \{Outdoor\}.
    \end{tcolorbox}
    
    \begin{tcolorbox}[width=\linewidth, colback=gray!5, colframe=gray, boxrule=0.5pt, title=Prompt for Clothing Items]
    Imagine you're creating metadata for an image database of clothing items. Your task is to select five keywords that best represent the image content. Fill in the blanks: \textbf{[Type]}, \textbf{[Color]}, \textbf{[Wear Location]}, \textbf{[Material]}, \textbf{[Style]}.\\[0.5em]
    For example, your answer will be: [Type] \{Dress\}, [Color] \{Red\}, [Wear Location] \{Torso\}, [Material] \{Cotton\}, [Style] \{Casual\}.
    \end{tcolorbox}
    \caption{Domain-specific prompts designed to guide LVLMs in generating structured keyword-based descriptions for items from the selected datasets.}
    \label{fig:lvlm_icl}
\end{figure*}

\begin{table}[!t]
\small
    \caption{Statistics of the datasets.}\label{tab:datasetInfo}
    \centering
    \begin{tabular}{lcccc}
    \toprule
        \textbf{Datasets} & $\bm{|\mathcal{U}|}$ & $\bm{|\mathcal{I}|}$ & $\bm{|\mathcal{R}|}$ & \textbf{Sparsity (\%)}\\ \cmidrule{1-5}
        Baby & 108,897 & 17,836 & 330,771 & 99.983\% \\
        Pets & 461,828 & 61,060 & 1,772,584 & 99.993\% \\ 
        Clothing & 398,796 & 75,616 & 2,109,975 & 99.993\% \\
        \bottomrule
    \end{tabular}
\end{table}

\subsection{Recommender Models}

Here are the models selected as baselines. We compare our approach against a broad set of baseline methods to provide a comprehensive performance landscape, spanning traditional collaborative filtering techniques, content-aware models, and recent multimodal recommendation frameworks. Within traditional collaborative filtering, we include foundational models such as \textbf{Item-kNN}~\cite{DBLP:conf/www/SarwarKKR01}, a memory-based approach that calculates item similarity based on user interaction history; \textbf{BPR-MF}~\cite{DBLP:journals/computer/KorenBV09}, a widely used matrix factorization model optimized for personalized ranking; and \textbf{LightGCN}~\cite{DBLP:conf/sigir/0001DWLZ020}, a simplified graph-based model focusing on essential neighborhood aggregation. To evaluate the impact of incorporating basic content features, we include \textbf{Attribute Item-kNN}~\cite{DBLP:conf/sigir/ScheinPUP02,5116421}, an extension of Item-kNN that utilizes item content, such as textual information, to compute similarity, which can be particularly beneficial for cold-start items. Representing various approaches to multimodal integration, we benchmark against several state-of-the-art multimodal recommender systems. \textbf{VBPR}~\cite{DBLP:conf/aaai/HeM16} is an early content-aware extension of BPR-MF that integrates pre-extracted visual features. \textbf{LATTICE}~\cite{DBLP:conf/mm/Zhang00WWW21} is a graph-based framework that learns enhanced item embeddings by mining latent structures and combining similarity graphs across different modalities using a graph convolutional network; these embeddings can then be used as input to standard latent factor models. \textbf{BM3}~\cite{DBLP:conf/www/ZhouZLZMWYJ23} is a self-supervised model that learns robust multimodal representations through contrastive learning, avoiding reliance on computationally expensive data augmentations. Finally, \textbf{FREEDOM}~\cite{DBLP:conf/mm/ZhouS23} is a multimodal model that improves recommendation by refining the user-item interaction graph using a frozen multimodal item-item similarity graph derived from LATTICE principles. This diverse selection allows us to rigorously investigate the influence of multimodal representations, isolate performance gains attributable to semantic content versus model capacity, and assess the efficacy of different multimodal integration strategies across various recommendation paradigms, directly addressing our research questions.
\begin{figure}[!b]
    \centering
    \includegraphics[width=1\linewidth]{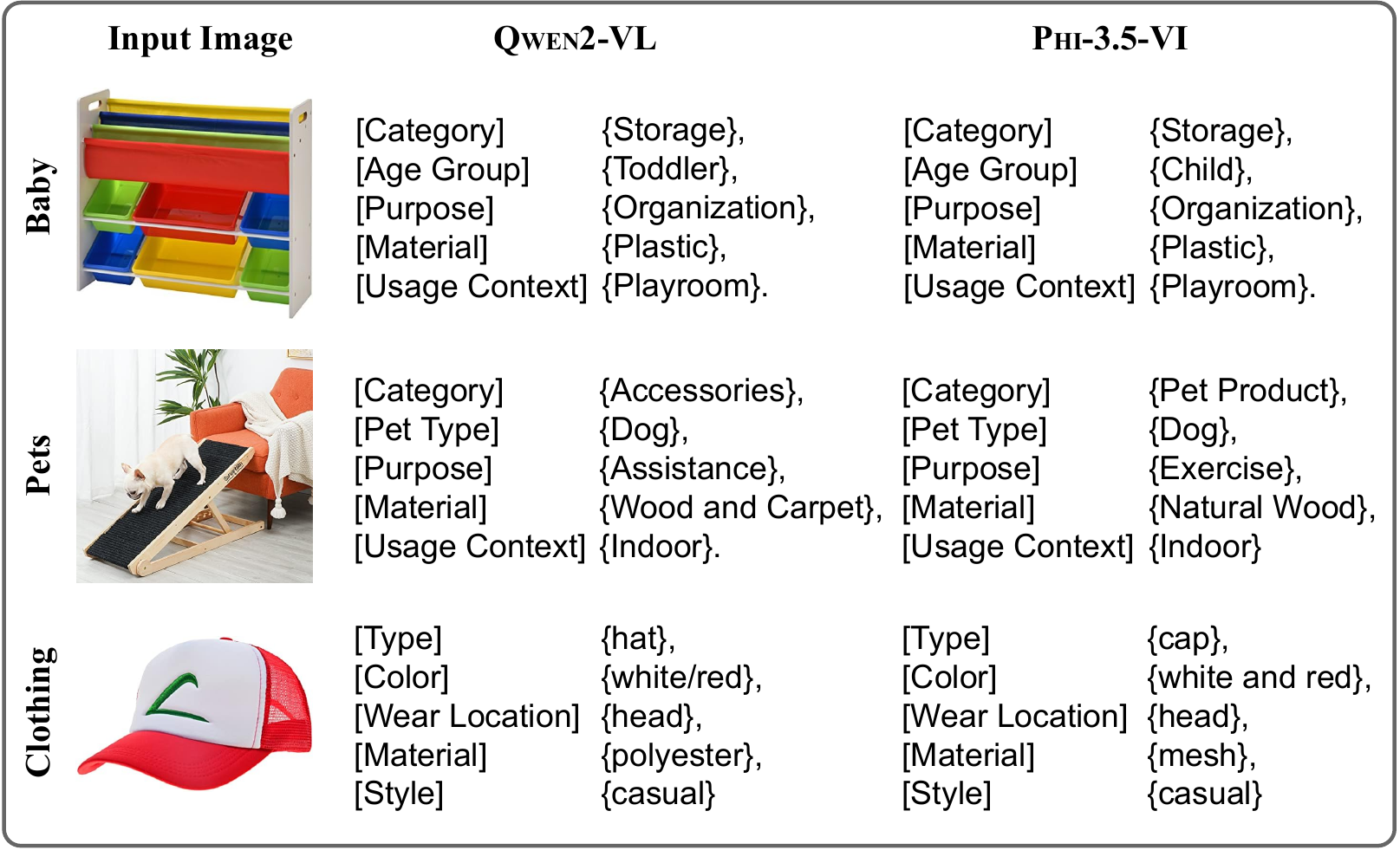}
    \caption{Examples of structured descriptions from LVLMs \qwen and \phiv for items in the Baby, Pets, and Clothing datasets.}
    \label{fig:lvlm_answers}
\end{figure}

\subsection{LVLMs Prompting and Answer Processing}\label{subsec:lvlm_preprocessing}
We employed the LVLMs~\qwen and~\phiv for their proven performance in multimodal tasks like VQA. Each item image was described using concise, domain-specific prompts designed to produce structured, essential outputs without excessive irrelevant detail.
For each of the three datasets, a tailored prompt format was defined based on visual inspection of item distributions. Each prompt included five representative keywords, a number selected for its balance between descriptiveness and discriminative power in qualitative evaluations. As shown in~\Cref{fig:lvlm_icl}, a single in-context example guided the generation process, enabling the models to adapt without fine-tuning. The keyword-based outputs adhered to the intended structure and reflected the LVLMs’ nuanced understanding of the specified attributes~(\Cref{fig:lvlm_answers}). For instance, on the Clothing dataset with \qwen, the process yielded 3,239 unique keywords.


\definecolor{lightblue}{RGB}{173,216,230} 
\definecolor{mediumblue}{RGB}{100,149,237} 
\definecolor{darkblue}{RGB}{0,0,139}       

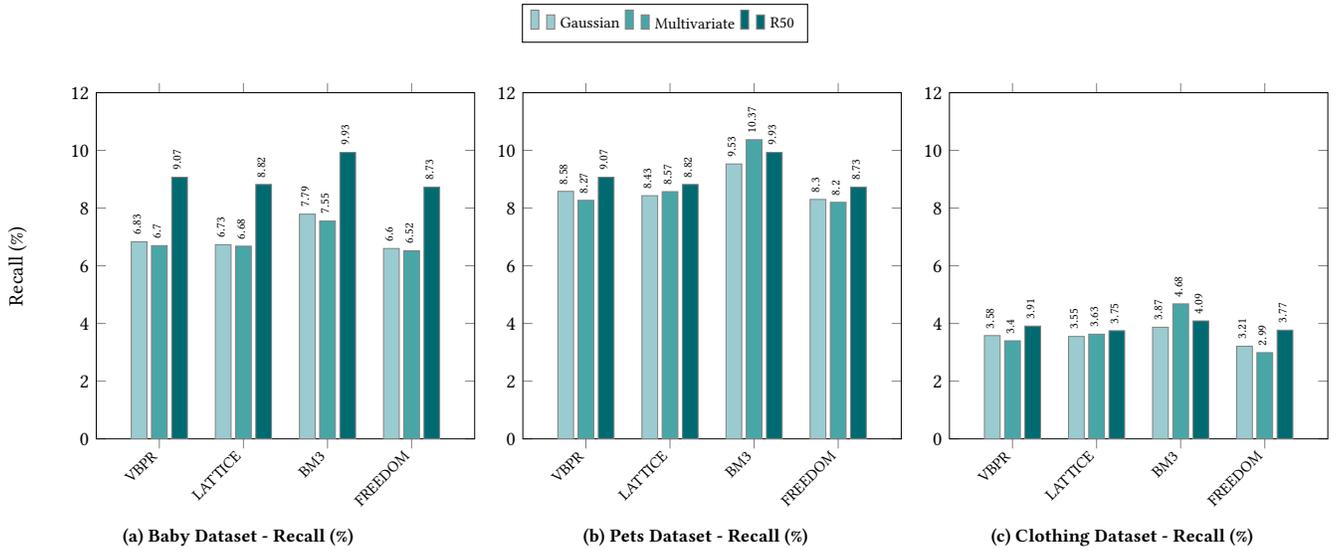
\begin{figure*}[!ht] 
\centering 

\begin{tikzpicture}
        \node[inner sep=0pt] (legendpos) {}; 
        \node at (legendpos.center) {\pgfplotslegendfromname{sharedlegend}};
    \end{tikzpicture}

\subfloat[Baby Dataset - Recall (\%)]{
\begin{tikzpicture}[scale=0.85] 
        \begin{axis}[
            width=7.5cm, 
            height=7cm, 
            ylabel={Recall (\%)},
            ybar,
            ymin=0,
            ymax=12, 
            symbolic x coords={VBPR, LATTICE, BM3, FREEDOM},
            xtick=data,
            x tick label style={rotate=45,anchor=east,font=\footnotesize},
            bar width=7pt, 
            enlarge x limits=0.25, 
            nodes near coords,
            every node near coord/.append style={
                font=\tiny,
                rotate=90,
                anchor=west,
                yshift=1pt 
            },
            nodes near coords style={/pgf/number format/fixed, /pgf/number format/precision=2} 
        ]
            \addplot[style = {fill=color1, draw=black!50}] coordinates {
                (VBPR, 6.83) (LATTICE, 6.73) (BM3, 7.79) (FREEDOM, 6.60) 
            };
            \addplot[style = {fill=color2, draw=black!50}] coordinates {
                (VBPR, 6.70) (LATTICE, 6.68) (BM3, 7.55) (FREEDOM, 6.52) 
            };
            \addplot[style = {fill=color3, draw=black!50}] coordinates {
                (VBPR, 9.07) (LATTICE, 8.82) (BM3, 9.93) (FREEDOM, 8.73) 
            };
        \end{axis}
    \end{tikzpicture}
}
\hfill 
\subfloat[Pets Dataset - Recall (\%)]{
    \begin{tikzpicture}[scale=0.85]
        \begin{axis}[
            width=7.5cm,
            height=7cm,
            ybar,
            ymin=0,
            ymax=12,
            symbolic x coords={VBPR, LATTICE, BM3, FREEDOM},
            xtick=data,
            x tick label style={rotate=45,anchor=east,font=\footnotesize},
            bar width=7pt,
            enlarge x limits=0.25,
            legend style={
                at={(0.5,-0.30)}, 
                anchor=north,
                legend columns=-1,
                font=\scriptsize
            },
            legend to name=sharedlegend,
            nodes near coords,
            every node near coord/.append style={
                font=\tiny,
                rotate=90,
                anchor=west,
                yshift=1pt
            },
            nodes near coords style={/pgf/number format/fixed, /pgf/number format/precision=2}
        ]
            \addplot[style = {fill=color1, draw=black!50}] coordinates {
                (VBPR, 8.58) (LATTICE, 8.43) (BM3, 9.53) (FREEDOM, 8.30) 
            };
            \addlegendentry{Gaussian}
            \addplot[style = {fill=color2, draw=black!50}] coordinates {
                (VBPR, 8.27) (LATTICE, 8.57) (BM3, 10.37) (FREEDOM, 8.20) 
            };
            \addlegendentry{Multivariate}
            \addplot[style = {fill=color3, draw=black!50}] coordinates {
                (VBPR, 9.07) (LATTICE, 8.82) (BM3, 9.93) (FREEDOM, 8.73) 
            };
            \addlegendentry{R50}
        \end{axis}
    \end{tikzpicture}
}
\hfill
\subfloat[Clothing Dataset - Recall (\%)]{
\begin{tikzpicture}[scale=0.85]
        \begin{axis}[
            width=7.5cm,
            height=7cm,
            ybar,
            ymin=0,
            ymax=12, 
            symbolic x coords={VBPR, LATTICE, BM3, FREEDOM},
            xtick=data,
            x tick label style={rotate=45,anchor=east,font=\footnotesize},
            bar width=7pt,
            enlarge x limits=0.25,
            nodes near coords,
            every node near coord/.append style={
                font=\tiny,
                rotate=90,
                anchor=west,
                yshift=1pt
            },
            nodes near coords style={/pgf/number format/fixed, /pgf/number format/precision=2}
        ]
            \addplot[style = {fill=color1, draw=black!50}] coordinates {
                (VBPR, 3.58) (LATTICE, 3.55) (BM3, 3.87) (FREEDOM, 3.21) 
            };
            \addplot[style = {fill=color2, draw=black!50}] coordinates {
                (VBPR, 3.40) (LATTICE, 3.63) (BM3, 4.68) (FREEDOM, 2.99) 
            };
            \addplot[style = {fill=color3, draw=black!50}] coordinates {
                (VBPR, 3.91) (LATTICE, 3.75) (BM3, 4.09) (FREEDOM, 3.77) 
            };
        \end{axis}
    \end{tikzpicture}
}
\caption{ Recall@20 (\%) performance for four MMRSs on (a) Baby, (b) Pets, and (c) Clothing datasets. Models are evaluated using item representations from: (i) Gaussian noise, (ii) multivariate structured noise, and (iii) \rnetext visual embeddings.}
\label{fig:recall_histograms_subplots}
\end{figure*}

To assess the utility of LVLMs in recommendation, we adopt a twofold evaluation strategy. First, we evaluate the impact of the latent representations extracted from LVLMs within MMRSs, specifically, the embeddings associated with the \texttt{[EOS]} token that encapsulates the generated response (see~\Cref{subsec:mm_feat_extractors}). Second, to assess the semantic informativeness and multimodal understanding embedded in the generated descriptions, and the corresponding \texttt{[EOS]} representations, we incorporate them as additional content signals into a classical hybrid model (e.g., Attribute Item-kNN). For comprehensive benchmarking, this approach is compared against both standard collaborative baselines and MMRSs that do not utilize such descriptive content. To enable this integration, raw textual descriptions are transformed into structured keyword-based features through minimal preprocessing. This involves correcting formatting inconsistencies, applying lemmatization, and clustering keywords based on semantic similarity. The top-50 most frequent keywords per category are retained, while less frequent terms are mapped to a generic “Other” token. The final representation consists of 255 one-hot encoded features, which mitigates sparsity and improves the robustness of the Attribute Item-kNN model.

\subsection{Implementation Details}
\label{subsec:implementation}
Feature extraction was performed using the \texttt{Ducho} framework~\cite{DBLP:conf/www/AttimonelliDMPG24}. We extracted visual features with the \rnet, \vit, \sbert, and CLIP backbones, and integrated LVLMs features \qwen and \phiv via HuggingFace\footnote{\url{https://huggingface.co}} models to obtain the \texttt{[EOS]} embeddings of the generated descriptions. These embeddings served as content representations in most recommendation models.

Recommendation training and evaluation were conducted using the \texttt{Elliot} framework~\cite{Elliot}. We tested both similarity-based and learning-based recommenders. For Item-kNN and Attribute Item-kNN, we evaluated multiple similarity functions: \texttt{cosine}, \texttt{jaccard}, \texttt{dot}, \texttt{asym}, and \texttt{tversky}. The number of neighbors was varied within the range $\{5, 100\}$ and different feature weighting schemes were evaluated, including \texttt{TF\_IDF} and \texttt{BM25}. In the specific case of Attribute Item-kNN, discrete keywords derived from the LVLMs text were required, as described in~\Cref{subsec:lvlm_preprocessing}. In particular, lemmatization was performed using the NLTK toolkit~\cite{NLTK}, keywords were grouped applying agglomerative clustering using cosine distance, via the Sentence Transformers library\footnote{\url{https://sbert.net}}.
For learning-based recommenders, we conducted a hyperparameter search across learning rates $\{0.0001, 0.0005, 0.001, 0.005, 0.01\}$, regularization values $\{10^{-2}, 10^{-1}\}$, and latent dimensions $\{64, 128, 256\}$, resulting in 30 explorations for each configuration.

Model performance was evaluated using standard top-$K$ recommendation metrics: Recall@20, nDCG@20, and Hit Ratio@20. The best hyperparameters for each model were selected based on Recall@20 from the validation set to ensure a fair comparison. All experiments were conducted on a compute node with an NVIDIA A100 GPU, involving over \textit{7,000} distinct training runs. The codebase for reproducing our experiments is available at \href{http://split.to/mmrs}{http://split.to/mmrs}.
\section{Results}

This section presents the results of the experimental evaluation, structured around the following research questions (RQs):
\begin{enumerate}[leftmargin=0.9cm]
\item[\textbf{RQ1:}] \textit{Do multimodal recommender systems truly benefit from the semantic content of input features, or are observed performance gains mainly relates to increased model complexity?}

\item[\textbf{RQ2:}] \textit{Is there a principled and embedding-agnostic way to provide multimodal content to recommender systems?}

\item[\textbf{RQ3:}] \textit{Do LVLMs embeddings faithfully capture semantic item content relevant to recommendation?}

\end{enumerate}

\subsection{Impact of Item Features in Recommendation (\textit{RQ1})}
An important question is whether MMRSs genuinely leverage the semantic information embedded in item representations to enrich the user-item interaction matrix, or the observed performance improvements are primarily attributable to the increased model capacity, such as a larger number of parameters, required to process additional input modalities. To address this, we evaluate four representative MMRS architectures: VBPR, LATTICE, BM3, and FREEDOM (see~\Cref{subsec:mm_feat_extractors}), under controlled input settings. Specifically, we examine model behavior when multimodal inputs are represented by: (i) pure Gaussian noise matching the dimensionality of standard item embeddings; (ii) multivariate noise designed to match the statistical properties (e.g., mean and variance) of real embeddings; and (iii) visual features extracted from~\rnet, a widely adopted feature extractor in multimodal recommendation.

\begin{table*}[!ht]
\caption{Top-20 performance (Recall, nDCG, HR as \%) of MMRS models using different feature extractors on Baby, Pets, and Clothing datasets. Boldface/\underline{underline} denote best/second-best extractor performance within each MMRSs model. $^{\star}$  indicates that the improvement of the best-performing extractor over the second-best is statistically significant ($p < 0.05$) within the specific set of evaluated extractors for that MMRSs model.
}\label{tab:benchmarking_rq2} 
\centering 
\begin{adjustbox}{width=0.8\textwidth, center} 
\begin{tabular}{llccccccccc}
\toprule
\multirow{2}{*}{\textbf{Models}} & \multirow{2}{*}{\textbf{Extractors}} & \multicolumn{3}{c}{\textbf{Baby}} & \multicolumn{3}{c}{\textbf{Pets}} & \multicolumn{3}{c}{\textbf{Clothing}} \\ \cmidrule(lr){3-5} \cmidrule(lr){6-8} \cmidrule(lr){9-11}
& & $\text{Recall}\uparrow$ & $\text{nDCG}\uparrow$ & $\text{HR}\uparrow$ & $\text{Recall}\uparrow$ & $\text{nDCG}\uparrow$ & $\text{HR}\uparrow$ & $\text{Recall}\uparrow$ & $\text{nDCG}\uparrow$ & $\text{HR}\uparrow$ \\ \cmidrule{1-11}

\multirow{6}{*}{VBPR} 
& \textsc{\qwen} & \textbf{7.690\rlap{$^\star$}} & \textbf{3.460\rlap{$^\star$}} & \textbf{8.140\rlap{$^\star$}} & \textbf{9.340\rlap{$^\star$}} & \textbf{4.320\rlap{$^\star$}} & \textbf{10.180\rlap{$^\star$}} & \textbf{4.060\rlap{$^\star$}} & \textbf{1.820\rlap{$^\star$}} & \textbf{4.810\rlap{$^\star$}} \\
& \textsc{\phiv} & \underline{7.640} & \underline{3.440} & \underline{8.100} & \underline{9.260} & \underline{4.280} & \underline{10.100} & \underline{3.990} & \underline{1.800} & \underline{4.720} \\
& \textsc{CLIP} & 7.490 & 3.340 & 7.940 & 9.170 & 4.220 & 10.010 & 3.960 & 1.780 & 4.690 \\
& \textsc{\rnet~--~\sbert} & 7.550 & 3.370 & 8.020 & 9.230 & 4.250 & 10.070 & 3.940 & 1.780 & 4.670 \\
& \textsc{\rnet} & 7.400 & 3.300 & 7.860 & 9.070 & 4.180 & 9.900 & 3.910 & 1.760 & 4.630 \\
& \textsc{\vit} & 7.390 & 3.270 & 7.840 & 8.990 & 4.140 & 9.810 & 3.940 & 1.780 & 4.660 \\ 
 \cmidrule{1-11}

\multirow{6}{*}{LATTICE} 
& \textsc{\qwen} & \underline{7.410} & \underline{3.210} & \underline{7.860} & \underline{9.030} & 4.080 & \underline{9.830} & \textbf{4.050\rlap{$^\star$}} & \textbf{1.800\rlap{$^\star$}} & \textbf{4.790\rlap{$^\star$}} \\
& \textsc{\phiv} & 7.330 & \underline{3.210} & 7.790 & \textbf{9.070\rlap{$^\star$}} & \textbf{4.110\rlap{$^\star$}} & \textbf{9.880\rlap{$^\star$}} & 3.850 & 1.710 & 4.560 \\
& \textsc{CLIP} & 7.150 & 3.130 & 7.610 & 8.740 & 3.930 & 9.530 & 3.771 & 1.680 & 4.480 \\
& \textsc{\rnet~--~\sbert} & \textbf{7.450} & \textbf{3.260\rlap{$^\star$}} & \textbf{7.890} & 9.020 & \underline{4.090} & \underline{9.830} & \underline{3.910} & \underline{1.750} & \underline{4.630} \\
& \textsc{\rnet} & 7.240 & 3.120 & 7.680 & 8.820 & 3.970 & 9.610 & 3.750 & 1.670 & 4.440 \\
& \textsc{\vit} & 7.200 & 3.130 & 7.630 & 8.820 & 3.990 & 9.640 & 3.780 & 1.680 & 4.480 \\
 \cmidrule{1-11}

\multirow{6}{*}{BM3}
& \textsc{\qwen} & 7.630 & 3.340 & 8.080 & \textbf{10.360} & \textbf{4.950} & \textbf{11.320} & \underline{4.680} & \underline{2.190} & \underline{5.570} \\
& \textsc{\phiv} & \textbf{7.680\rlap{$^\star$}} & \textbf{3.370} & \textbf{8.130\rlap{$^\star$}} & \underline{10.290} & \underline{4.920} & \underline{11.250} & \textbf{4.700\rlap{$^\star$}} & \textbf{2.210\rlap{$^\star$}} & \textbf{5.610\rlap{$^\star$}} \\
& \textsc{CLIP} & \underline{7.670} & \underline{3.350} & \textbf{8.130\rlap{$^\star$}} & 9.510 & 4.340 & 10.370 & 3.990 & 1.790 & 4.750 \\
& \textsc{\rnet~--~\sbert} & 7.660 & 3.290 & \underline{8.100} & 9.510 & 4.330 & 10.370 & 3.950 & 1.760 & 4.690 \\
& \textsc{\rnet} & 7.640 & 3.290 & 8.080 & 9.930 & 4.600 & 10.850 & 4.090 & 1.860 & 4.870 \\
& \textsc{\vit} & 7.600 & 3.320 & 8.050 & 9.700 & 4.470 & 10.580 & 4.150 & 1.900 & 4.930 \\
 \cmidrule{1-11}

\multirow{6}{*}{FREEDOM} 
& \textsc{\qwen} & \underline{7.780} & \underline{3.400} & \underline{8.240} & \underline{9.120} & \underline{4.070} & \underline{9.900} & \underline{4.130} & \underline{1.810} & \underline{4.880} \\
& \textsc{\phiv} & 7.530 & 3.220 & 7.980 & 8.950 & 3.980 & 9.740 & 4.020 & 1.750 & 4.760 \\
& \textsc{CLIP} & 6.230 & 2.520 & 6.620 & 7.740 & 3.380 & 8.430 & 2.910 & 1.210 & 3.450 \\
& \textsc{\rnet~--~\sbert} & \textbf{7.810} & \textbf{3.420} & \textbf{8.270} & \textbf{9.470\rlap{$^\star$}} & \textbf{4.320\rlap{$^\star$}} & \textbf{10.300\rlap{$^\star$}}  & \textbf{4.220\rlap{$^\star$}} & \textbf{1.870\rlap{$^\star$}} & \textbf{4.990\rlap{$^\star$}} \\
& \textsc{\rnet} & 7.470 & 3.200 & 7.920 & 8.730 & 3.890 & 9.510 & 3.770 & 1.630 & 4.470 \\
& \textsc{\vit} & 7.420 & 3.190 & 7.860 & 8.840 & 3.940 & 9.620 & 3.900 & 1.700 & 4.620 \\
 \cmidrule{1-11}


\end{tabular} 
\end{adjustbox}
\end{table*}

This experimental setup allows us to isolate the effect of model capacity, as all MMRSs receive input embeddings of identical dimensionality. As shown in~\Cref{fig:recall_histograms_subplots}, which reports Recall@20 scores across the three datasets, VBPR, LATTICE, and FREEDOM consistently achieve higher performance when provided with semantically meaningful features (e.g., \rnet representations). This indicates that these models benefit from the informative content of the input features, rather than from increased capacity alone. 
Interestingly, BM3 performs similarly on the Pets and Clothing datasets regardless of input type, with a slight improvement on Baby when using \rnet features. While it does benefit from semantic features, the performance gap between noise-based and semantic inputs remains relatively small compared to other models, particularly on Pets and Clothing. This indicates that BM3's effectiveness may stem more from its architecture than in multimodal content. Its ability to utilize minimal structural cues in complex inputs might reduce its reliance on rich semantic features after a certain complexity threshold is reached.


With regard to \textbf{RQ1:} ``\textbf{Do multimodal recommender systems truly benefit from the semantic content of input features, or are observed performance gains mainly relates to increased model capacity?}'', our evaluation shows that MMRSs do indeed benefit from semantically meaningful item representations. This positive effect is apparent across most models; however, it is less pronounced for the BM3 model. This suggests that the performance improvements seen with BM3 may be more influenced by its architectural complexity rather than the quality of the input features.

\subsection{LVLMs Representations for Multimodal Recommendation (\textit{RQ2})}
Previous sections have demonstrated that multimodal representations can notably improve the performance of RSs. A widely adopted approach involves combining modality-specific features, typically extracted using dedicated encoders such as \rnet for visual inputs and \sbert for textual data.
However, this late-fusion approach has limitations, including potential semantic misalignment and limited control over the distribution of modality-specific information within the fused latent representation.
To evaluate these methods, we analyze various multimodal architectures in the recommendation pipeline, including: (i) \rnet and \vit as traditional vision backbones; (ii) a baseline that concatenates embeddings from \rnet and \sbert, in line with previous works~\cite{DBLP:journals/tors/MalitestaCPMNS25, DBLP:conf/mm/ZhouS23, DBLP:conf/mm/Zhang00WWW21, DBLP:conf/www/ZhouZLZMWYJ23}; and (iii) 
CLIP, a multimodal backbone. Since it is not trained on the Amazon Reviews 2023 dataset, its visual and textual representations may not align semantically within the shared latent space. Therefore, consistent with (ii), we concatenate CLIP's visual and textual embeddings to preserve the informational content of both modalities.
Additionally, we examine \qwen and \phiv for their ability to produce multimodal representations without fusion. These models use structured queries to generate concise and content-rich descriptions of item images~\cite{DBLP:journals/corr/abs-2404-14219, DBLP:journals/corr/abs-2409-12191, DBLP:journals/corr/abs-2409-02813}, providing unified embeddings as direct inputs for recommendation models.

Our empirical evaluation, presented in Table~\ref{tab:benchmarking_rq2}, shows that across all tested recommender models (VBPR, LATTICE, BM3, FREEDOM), LVLMs embeddings consistently rank among the top two performers.
A critical aspect of our analysis is the statistical significance of performance differences between the leading and runner-up extractors for each model. For instance, with the VBPR model, \qwen not only achieves the highest scores across all datasets (Baby, Pets, Clothing) and metrics (Recall, nDCG, HR), but these scores are also statistically significantly superior (p < 0.05, indicated by~$^{\star}$~in the table) to those of the second-best extractor, \phiv. This highlights that even within the LVLMs family, specific model choices can yield significant performance variations. Similarly, for the BM3 model, the best-performing LVLM extractor (\qwen on Pets; \phiv on Baby and Clothing) notably outperforms the respective second-best option in each case.

Although LVLMs demonstrate strong, often superior, performance, the traditional \rnet~--~\sbert combination remains a robust baseline in certain contexts. Notably, for the FREEDOM model, \rnet~--~\sbert consistently delivers the best results across all datasets, surpassing LVLMs extractors which hold the second position. 
In the LATTICE model, \rnet~--~\sbert performs better on Baby than the second-best LVLM, \qwen. 
However, LVLMs (\phiv on Pets and \qwen on Clothing) hold the significantly leading positions on the other datasets for LATTICE. These cases show that the performance gap between top and succeeding extractors is often statistically significant, highlighting the need for careful extractor selection based on the recommendation model and dataset characteristics.



\begin{table}[t!]
  \centering
  \caption{Aggregated Borda count scores and overall rank for feature extractors. Scores are derived from Recall@20 and general performance ranks across all MMRSs models and datasets, with higher scores indicating superior overall extractor performance.}
  \label{tab:borda_summary_recall}
  \begin{adjustbox}{width=0.4\textwidth, center}
   \begin{tabular}{cl r r r r} 
    \toprule
    Rank & Extractor   & Baby & Pets & Clothing & Overall \\
    \midrule
    1    & \textsc{\qwen}     & 14.0 & 18.0 & 18.0     & 50.0         \\
    2    & \textsc{\phiv}      & 15.0 & 16.0 & 15.0     & 46.0         \\
    3    & \textsc{\rnet~--~\sbert} & 16.0 & 11.5 & 10.5     & 38.0         \\
    4    & \textsc{\rnet}     & 7.0  & 6.5  & 3.0      & 16.5         \\
    5    & \textsc{\vit}          & 2.0  & 5.5  & 8.5      & 16.0         \\
    6    & \textsc{CLIP}         & 6.0  & 2.5  & 5.0      & 13.5         \\
    \bottomrule
  \end{tabular}
  \label{tab:borda_count}
  \end{adjustbox}
\end{table}

\begin{table*}[!ht]
\caption{Top-20 performance (Recall, nDCG, HR as \%) for all configurations of RSs, extractors, and datasets. Boldface/\underline{underlined} denote best/second-best values for each recommendation model. $^{\star}$ indicates $p < 0.05$ significance.
}\label{tab:benchmarking_rq3} 
\centering 
\begin{adjustbox}{width=0.8\textwidth, center} 
\begin{tabular}{llccccccccc}
\toprule
\multirow{2}{*}{{Models}} & \multirow{2}{*}{{Extractors}} & \multicolumn{3}{c}{{Baby}} & \multicolumn{3}{c}{{Pets}} & \multicolumn{3}{c}{{Clothing}} \\ \cmidrule(lr){3-5} \cmidrule(lr){6-8} \cmidrule(lr){9-11}
& & $\text{Recall}\uparrow$ & $\text{nDCG}\uparrow$ & $\text{HR}\uparrow$ & $\text{Recall}\uparrow$ & $\text{nDCG}\uparrow$ & $\text{HR}\uparrow$ & $\text{Recall}\uparrow$ & $\text{nDCG}\uparrow$ & $\text{HR}\uparrow$ \\ \cmidrule{1-11}

Random & \textsc{---} & 1.046 & 0.362 & 0.115 & 0.036 & 0.014 & 0.039 & 0.028 & 0.010 & 0.033 \\
MostPop & \textsc{---} & 5.968 & 2.358 & 4.173 & 4.173 & 1.780 & 4.512 & 1.971 & 0.081 & 2.253 \\
Item-kNN & --- & 6.785 & 3.159 & 7.233 & 9.229 & 4.523 & 10.061 & 3.997 & 1.973 & 4.786 \\
BPR-MF & --- & 7.031 & 2.903 & 7.440 & 8.787 & 4.024 & 9.610 & 3.572 & 1.595 & 4.255 \\ 
LightGCN & --- & 7.468 & 3.168 & 7.918 & 9.156 & 4.122 & 10.003 & 3.801 &1.638 & 4.500 \\ \cmidrule{1-11}

\multirow{2}{*}{Attribute Item-kNN}
& \textsc{Qwen2-VL} & 6.965 & 3.256 & 7.401 & 9.357 & 4.598 & 10.195 & 4.073 & 2.012 & 4.860 \\ 
& \textsc{\phiv} & 6.980 & 3.260 & 7.410 & 9.370 & 4.600 & 10.200 & 4.070 & 2.010 & 4.860 \\ \cmidrule{1-11}

\multirow{2}{*}{VBPR} 
& \textsc{\qwen} & \underline{7.690} & \textbf{3.460\rlap{$^\star$}} & \underline{8.140} & 9.340 & 4.320 & 10.180 & 4.060 & 1.820 & 4.810 \\
& \textsc{\phiv} & 7.640 & \underline{3.440} & 8.100 & 9.260 & 4.280 & 10.100 & 3.990 & 1.800 & 4.720 \\
 \cmidrule{1-11}

\multirow{2}{*}{LATTICE} 
& \textsc{\qwen} & 7.410 & 3.210 & 7.860 & 9.030 & 4.080 & 9.830 & 4.050 & 1.800 & 4.790 \\
& \textsc{\phiv} & 7.330 & 3.210 & 7.790 & 9.070 & 4.110 & 9.880 & 3.850 & 1.710 & 4.560 \\
 \cmidrule{1-11}

\multirow{2}{*}{BM3}
& \textsc{\qwen} & 7.630 & 3.340 & 8.080 & \textbf{10.360} & \textbf{4.950} & \textbf{11.320} & \underline{4.680} & \underline{2.190} & \underline{5.570} \\
& \textsc{\phiv} & 7.680 & 3.370 & 8.130 & \underline{10.290} & \underline{4.920} & \underline{11.250} & \textbf{4.700\rlap{$^\star$}} & \textbf{2.210\rlap{$^\star$}} & \textbf{5.610\rlap{$^\star$}} \\
 \cmidrule{1-11}

\multirow{2}{*}{FREEDOM} 
& \textsc{\qwen} & \textbf{7.780\rlap{$^\star$}} & 3.400 & \textbf{8.240\rlap{$^\star$}} & 9.120 & 4.070 & 9.900 & 4.130 & 1.810 & 4.880 \\
& \textsc{\phiv} & 7.530 & 3.220 & 7.980 & 8.950 & 3.980 & 9.740 & 4.020 & 1.750 & 4.760 \\
 \cmidrule{1-11}


\end{tabular} 
\end{adjustbox}
\end{table*}

Despite these model-specific nuances where traditional methods can occasionally excel, a broader perspective is offered by the Borda count analysis (Table~\ref{tab:borda_count}).
This analysis provides an aggregated view of extractor performance. In essence, for each specific evaluation scenario (a given recommender model, dataset, and metric), the extractors are ranked. Points are awarded based on these ranks (e.g., the top-performing extractor gets the most points, the second gets fewer, and so on). These points are then summed across all scenarios for each extractor. The resulting total score offers a consolidated measure of which extractor performs the most consistently well across the diverse range of tested conditions, helping to identify generally superior approaches beyond isolated best-case results.
This aggregate view reveals that, on average, \qwen and \phiv secure the highest rankings across the majority of extractor-dataset-model-metric combinations. This evidence strongly supports the efficacy of LVLMs embeddings, attributing their success to the creation of semantically grounded, multimodal-by-design representations. 

In response to \textbf{RQ2:} ``\textbf{Is there a principled and embedding-agnostic way to provide multimodal content to recommender systems?}'', the answer is yes, using LVLMs. Experimental results indicate that multimodal-by-design embeddings associated with LVLMs generated descriptions consistently outperform classical fusion approaches, such as the combination of \rnet and \sbert, as well as alternative multimodal-by-design solutions like CLIP. 
These embeddings also outperform unimodal visual backbones such as \rnet and \vit, demonstrating the semantic richness of LVLMs representations. By avoiding late-fusion of independently derived modality-specific embeddings, LVLMs representations enable a more coherent and controlled integration of multimodal signals, reducing the risk of cross-modal misalignment.

\subsection{Semantic Alignment of LVLMs Representations (\textit{RQ3})}\label{subsec:rq3}
As previously discussed, multimodal-by-design representations generated by LVLMs frequently improve the performance of MMRSs compared to other multimodal configurations. In particular, the benefits extend beyond improvements in embedding quality alone. As detailed in~\Cref{sec:methodology,subsec:lvlm_preprocessing}, these embeddings are derived from the final hidden state corresponding to the \texttt{[EOS]} token in the LVLMs output. By exploiting the autoregressive nature of LVLMs, this token encapsulates the complete semantic content of the generated textual description, effectively representing its multimodal understanding. Consequently, each LVLM embedding is intrinsically linked to a coherent, human-readable textual description generated by the model itself.


To evaluate the semantic richness and utility of the generated descriptions, we incorporate them as additional content features into the classical hybrid model Attribute Item-kNN. As detailed in~\Cref{subsec:lvlm_preprocessing}, LVLMs generated descriptions are pre-processed into keyword-based features and converted into one-hot vectors for seamless integration. We then assess the impact of these features on recommendation performance relative to both traditional collaborative filtering baselines (Random, MostPop, Item-kNN, BPR-MF, LightGCN) and multimodal recommendation systems (VBPR, LATTICE, BM3, FREEDOM) that directly utilize LVLMs embeddings.


The comparative results presented in~\Cref{tab:benchmarking_rq3} demonstrate that incorporating LVLMs textual descriptions into Attribute Item-kNN yields notable performance improvements compared to all baseline methods. While these gains are moderate on the Baby dataset, they are particularly pronounced on the Pets and Clothing datasets, with the exception of BM3 on Pets which retains a performance advantage when utilizing LVLMs embeddings. In general, Attribute Item-kNN, especially when leveraging \phiv attributes, not only surpasses all classical baselines but also matches or exceeds the performance of several advanced MMRSs. For example, on the Clothing dataset, Attribute Item-kNN consistently outperforms models such as VBPR and LATTICE. 
Overall, Attribute Item-kNN demonstrates strong competitiveness, with consistently high nDCG scores that highlight the notable ranking capabilities of kNN-based models. This effect is particularly evident when the model is enhanced with LVLMs content, emphasizing the considerable informativeness and practical utility of these multimodal textual descriptions.

In light of previous observations, we positively address \textbf{RQ3:} “\textbf{Do LVLMs embeddings faithfully capture semantic item content relevant to recommendation?}”. By extracting keywords from LVLMs textual descriptions and incorporating them as features into a straightforward content-based recommender, Attribute Item-kNN, we demonstrate consistent performance improvements. These gains, observed across multiple datasets, frequently allow this content-aware approach to outperform classical collaborative filtering methods and, in several cases, advanced MMRSs that directly leverage LVLMs embeddings. This provides strong quantitative evidence of the semantic informativeness and practical utility of LVLMs textual content. The dual use of LVLMs outputs, as both latent embeddings and effective textual features, enhances even simple models, elevating their performance to competitive, and often statistically superior, levels. These findings underscore the value of LVLMs in generating rich multimodal representations for recommendation


\section{Conclusion}
This research demonstrates that the effectiveness of MMRSs is primarily influenced by the semantic content of multimodal inputs rather than solely by the increased model complexity. Our study shows that LVLMs provide a robust and systematic approach to create items embeddings \textit{multimodal-by-design}. These embeddings effectively capture cross-modal semantics without requiring explicit fusion, consistently improving recommendation performance across various MMRS architectures and often outperforming traditional late-fusion techniques.
A key contribution of this work is the validation of the semantic relevance of the representations generated by LVLMs. The ability to convert these embeddings into structured textual descriptions enhances classical recommender models when used as supplementary content. This provides direct evidence of the models' deep understanding of multimodal data and their practical usefulness. The dual capability of offering both powerful latent representations and interpretable textual outputs underscores the value of LVLMs.
Our findings support the adoption of high-quality, semantically aligned multimodal representations as a fundamental element in advancing recommendation systems. LVLMs present a promising foundation for developing more powerful, interpretable, and effective MMRSs. Future research may focus on tailoring LVLMs specifically for recommendation systems, exploring two primary aspects: first, utilizing LVLMs directly to generate recommendations, and second, enhancing the quality of explanations provided alongside those recommendations.
\newpage



\bibliographystyle{ACM-Reference-Format}
\bibliography{bibliography}

\end{document}